\documentclass[9pt]{article}
\usepackage{spconf,amsmath,graphicx}

\usepackage[]{xcolor}
\usepackage{caption}
\usepackage{subcaption}
\usepackage{booktabs}
\usepackage{float}
\usepackage{hyperref}
\usepackage{multirow}
\usepackage{amssymb}
\title{VoxWatch: An open-set speaker recognition benchmark on VoxCeleb}
\name{Raghuveer Peri, Seyed Omid Sadjadi, Daniel Garcia-Romero}
\address{AWS AI Labs}
\begin{document}

\maketitle
 
\begin{abstract}
Despite its broad practical applications such as in fraud prevention, open-set speaker identification (OSI) has received  less attention in the speaker recognition community compared to speaker verification (SV). OSI deals with determining if a test speech sample belongs to a speaker from a set of pre-enrolled individuals (in-set) or if it is from an out-of-set speaker. In addition to the typical challenges associated with speech variability, OSI is prone to the ``false-alarm problem''; as the size of the in-set speaker population (a.k.a watchlist) grows, the out-of-set scores become larger, leading to increased false alarm rates. This is in particular challenging for applications in financial institutions and border security where the watchlist size is typically of the order of several thousand speakers. 
Therefore, it is important to systematically quantify the false-alarm problem, and develop techniques that alleviate the impact of watchlist size on detection performance. Prior studies on this problem are sparse, and lack a common benchmark for systematic evaluations.
In this paper, we present the first public benchmark for OSI, developed using the VoxCeleb dataset. 
We quantify the effect of the watchlist size and speech duration on the watchlist-based speaker detection task using three strong neural network based systems.
In contrast to the findings from prior research, we show that the commonly adopted adaptive score normalization is not guaranteed to improve the performance for this task. 
On the other hand, we show that score calibration and score fusion, two other commonly used techniques in SV, result in significant improvements in OSI performance.
\end{abstract}

\section{Introduction}
\label{sec:intro}

Automatic speaker recognition has received much attention in recent times owing to its ubiquitous applications in telephone banking, customer service, virtual assistants, and smart appliances, to mention a few. In particular, speaker verification (SV), which deals with determining whether a speech sample belongs to the person claiming the identity, has been explored to a large extent. 
This has been facilitated by several open benchmarking evaluations such as the National Institute of Standards and Technology Speaker Recognition Evaluations (NIST SRE) \cite{sadjadi20192018} and VoxCeleb speaker recognition challenges (VoxSRC) \cite{chung2019voxsrc}.
On the other hand, open-set speaker identification (OSI) which is a generalization of SV and deals with determining whether a speech sample belongs to an individual from a set of pre-enrolled speakers, has remained relatively under-explored.
\begin{figure}[!t]
     \centering
      \includegraphics[scale=0.45]{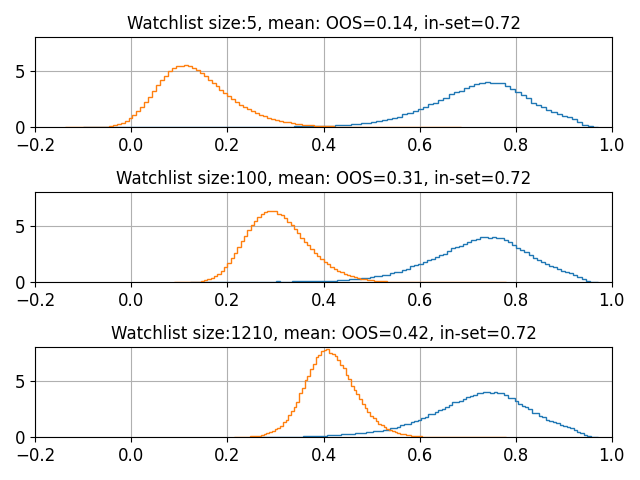}
      \caption{Cosine similarity score histograms for 3 different watchlist sizes. As the watchlist size grows, the out-of-set scores shift to the right, leading to higher false alarm rates.}
      \label{fig:ws_size_score}
      \vspace{-0.7cm}
\end{figure}
OSI has wide applications such as known fraudster detection in telephone banking services, group access authorization in smart home devices, audio based triaging, and speaker-based information retrieval from audio archives. In particular, financial institutions often maintain a list of known fraudsters (a.k.a a ``watchlist'') identified manually through suspicious activity detection or automatically via call/number spoofing detection \cite{ramasubramanian2012speaker}. 
Enrollment speech from these fraudsters is stored in the watchlist for comparison against incoming speech samples (from unknown speakers) to identify and prevent recidivism via negative recognition.
Despite its diverse applications, made more relevant by the recent rise in fraudulent activity, much of the prior work on OSI has focused on traditional speaker modeling techniques such as Gaussian mixture models (GMM) \cite{reynolds2000speaker} and i-vectors\cite{dehak2010front}. 
Modern neural network based methods for speaker modeling have thus far received little attention for OSI applications.

In addition to the challenges known for speaker recognition (e.g., intrinsic and extrinsic variabilities) \cite{hansen2018issues}, OSI suffers from the ``false-alarm problem" which stems from a shift in the distribution of non-target scores (test sample not from any of the watchlist speakers) as a function of the number of enrolled speakers (i.e., the size of the watchlist) \cite{ramasubramanian2012speaker,singer2004multi}. As shown in Figure~\ref{fig:ws_size_score}, a larger watchlist size results in the non-target or out-of-set (OOS) similarity scores shifting towards the right, leading to a larger overlap with the target (in-set) scores. 
This leads to a higher false alarm rate at a given fixed threshold. 
Intuitively, as the watchlist size grows, it is more likely for an OOS speaker to sound like one of the watchlist speakers, resulting in increased false alarms. Such degradation in performance can hinder the large-scale adoption of this technology, as typically such speaker spotting services are expected to work with watchlist sizes of thousands or more enrolled individuals. 
For comparison, in Iris-based biometrics applications where watchlist sizes of more than a million subjects are typical, negative recognition is performed with a very high accuracy at a relatively low false alarm rate \cite{ramasubramanian2012speaker}. 
Therefore, it is crucial to characterize the performance degradation due to the wacthlist size and develop techniques to improve this trade-off in OSI applications. 

Although this problem has been previously identified in the speaker recognition community \cite{ramasubramanian2012speaker,singer2004multi}, 
prior work that studied and tackled this issue is sparse, likely due to the lack of a standard publicly available benchmark. In this study, we aim to address this by creating a reproducible watchlist-based speaker detection benchmark using the VoxCeleb dataset \cite{nagrani2017voxceleb}.
Using this benchmark, we quantify the effect of watchlist size and speech duration on speaker detection performance. We evaluate a recent publicly available ResNet model as well as two state-of-the-art models trained in-house on the VoxCeleb dataset, demonstrating that stronger speaker discrimination leads to improved watchlist-based detection performance.
However, there is a limit by which speaker discrimination can be improved, especially using fixed amount of labelled data. Therefore, we also evaluate the effectiveness of the commonly adopted adaptive score normalization (AS-Norm) as well as quality measure based score calibration.

\vspace{-0.3cm}

\section{Related work}

A few prior studies have explored OSI, mostly using traditional speaker modeling techniques~\cite{ramasubramanian2012speaker,singer2004multi,gao2011osi,khoury2019pindrop,font2029mce,Wilkinghoff2020}. Ramasubramanian \cite{ramasubramanian2012speaker} presented a detailed study of OSI (called speaker spotting in the paper), its formulation, challenges and applications. They explored the effect of watchlist size on detection performance.
The analyses presented in these studies were limited to GMM-based systems on relatively small datasets. It remains unclear how modern neural network based systems perform on similar tasks.
A recent work used a time-delay neural network (TDNN) model trained on VoxCeleb and evaluated watchlist detection task using maximum watchlist size of $30$, which is too small for general OSI applications \cite{trnka2021speaker}.
Shon et al. \cite{Shon2019} presented the multi-target speaker detection and identification challenge evaluation (MCE2018) designed specifically to address the OSI task. However, audio recordings were not provided to the participants, and only i-vectors extracted from customer-agent conversations were shared. As noted by the challenge organizers, this was a limiting factor in exploring techniques that could potentially provide larger performance improvements.
Zigel and Wasserblat \cite{zigel2006deal} experimented with various score normalization techniques using a GMM-based system on watchlists of sizes 1 to 183, and observed that top-norm (using the top scoring trials as cohorts) performs the best. The top performing submission in the MCE2018 challenge \cite{khoury2019pindrop} used a fusion of multiple systems, and further showed the benefits of score normalization.
However, it remains unclear if the trends presented in these studies hold true for the recent neural network based systems, which are significantly better at representing speaker information than the traditional methods considered so far.
\vspace{-0.2cm}

\section{Dataset}
We conduct experiments using the VoxCeleb corpus \cite{nagrani2017voxceleb} to facilitate reproducible evaluations and encourage future research on OSI. We also publicly release the trial and speaker lists used in our experiments to allow for an open benchmarking setup for fair comparisons between this work and future research studies.
\vspace{-0.4cm}
\subsection{VoxCeleb}
The VoxCeleb corpus contains speech recordings from real-world, unconstrained interviews of several thousand celebrities. It consists of two separate datasets, VoxCeleb1 and VoxCeleb2, each of which is further split into development and test partitions. Henceforth, we refer to these splits as Vox1-dev~(1211 speakers), Vox1-test~(40 speakers), Vox2-dev~(5994 speakers) and Vox2-test~(118 speakers). The dataset contains speech segments split into utterances, with multiple utterances coming from the same video recording. 
In our experiments, we concatenate the segments from each video to form longer recordings to prevent the systems from potentially leveraging the correlations between speaker and channel/video information, as well as to enable embedding extraction from longer duration audio (e.g., 30s--90s).
\vspace{-0.2cm}

\begin{table}[]
\centering
\caption{Statistics of the trials used to evaluate watchlist detection performance. \# OOS trials decreases as watchlist size increases due to fewer OOS speakers.}
\resizebox{0.75\columnwidth}{!}{%
\begin{tabular}{cll}
\toprule
\textbf{Watchlist size} & \textbf{\# in-set trials} & \textbf{\# OOS trials} \\ \midrule
5              & 120,034                & 4,000,018           \\
10             & 120,020                & 4,000,018           \\
20             & 120,083                & 4,000,014           \\
50             & 118,931                & 2,763,061           \\
100            & 119,064                & 1,321,932           \\
200            & 118,888                & 601,610             \\
500           & 98,967                 & 141,199             \\
1210          & 4,001,144              & 120,083             \\ \bottomrule
\vspace{-1.cm}
\end{tabular}
\label{tab:stats}
}
\end{table}

\subsection{Watchlist generation}
We create watchlists of different sizes from the $1211$ speakers in Vox1-dev using an approach similar to k-fold cross-validation, where random groups~(of different sizes) of the speakers are chosen without replacement. 
This ensures that for a given watchlist size, each speaker is part of the in-set~(watchlist) speakers in exactly one watchlist.

For example, for a watchlist of size $50$, we create $24(=1211/50)$ separate speaker splits, and for each split the remaining $1161(=1211-50)$ speakers form the OOS speaker list. 
For a watchlist size of $1210$, 
we use a leave-one-speaker-out~(LOSO) approach, where $1211$ slightly different speaker splits are created containing $1210$ watchlist speakers. In this case each watchlist differs from the other by just $1$ speaker. Table \ref{tab:stats} shows the statistics of the trials used in our experiments for different watchlist sizes. For the watchlist size of $1210$, the number of in-set trials is large (with repetitions) due to the LOSO approach, and we sub-sample $\sim4M$ trials.

We create watchlists of sizes $5$, $10$, $20$, $50$, $100$, $200$, $500$ and $1210$ to cover the spectrum of applications for the speaker spotting task.
For example, a watchlist of size $5$ is typical in a smart home setting where the family can be imagined to be part of the in-set speakers with privileged access to security, online shopping, or entertainment features. 
Any person from outside the family (e.g., guests) are part of the OOS speakers with limited privileges. On the other hand, a watchlist size of a few hundreds is typical in a telephone banking service \cite{ramasubramanian2012speaker}.

\section{Task definitions}
\label{sec:task}

\begin{figure}[!t]
  \centering
  \includegraphics[width=\linewidth]{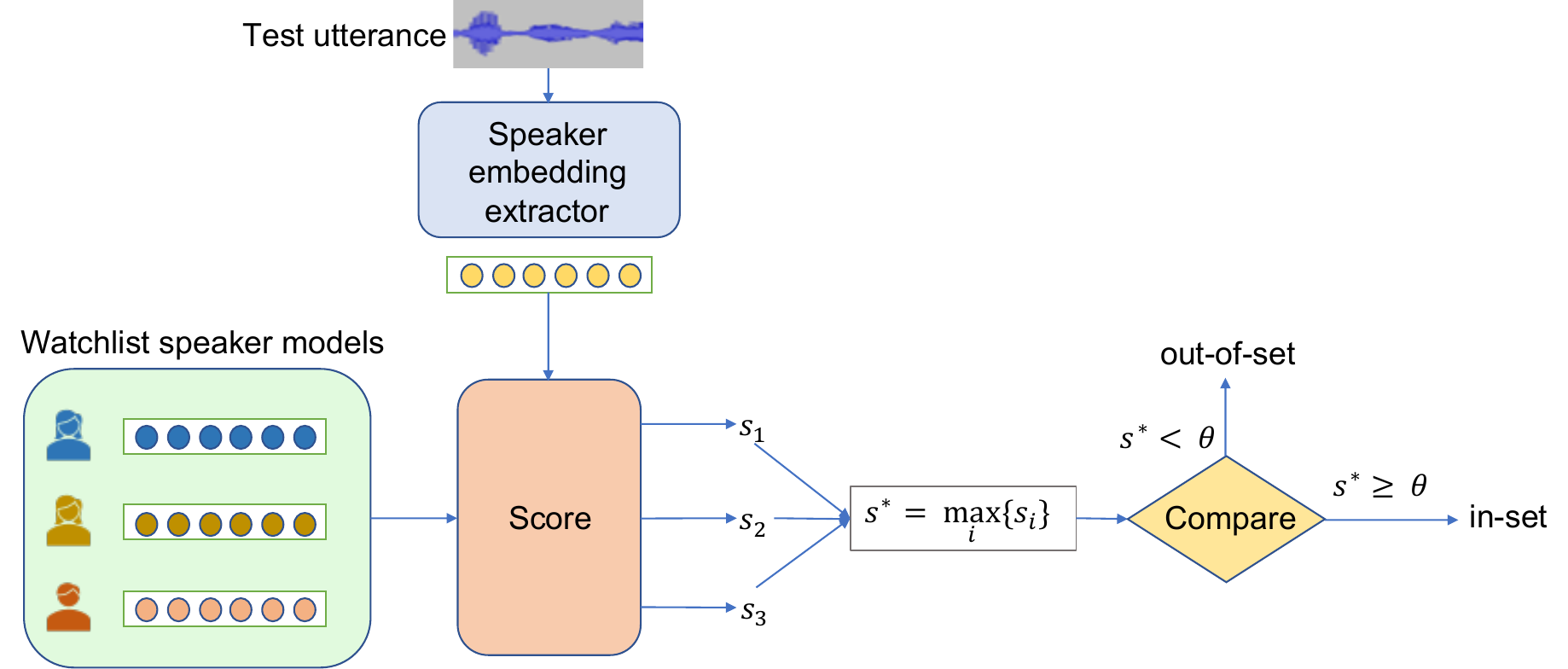}
  \caption{Block diagram showing the steps in watchlist-based speaker detection.}
  \vspace{-0.5cm}
  \label{fig:block}
\end{figure}

We pose the watchlist-based speaker detection task as a group detection problem following previous work \cite{ramasubramanian2012speaker, Shon2019}, and assume that each speaker in the watchlist has a pre-determined enrollment speech sample. As shown in Figure \ref{fig:block}, an incoming test sample from an unknown speaker is scored against all the enrolled in-set speakers to produce scores. As shown in (\ref{eq:1}), the maximum score is compared against a threshold, and the speech sample is labelled as originating from a watchlist speaker if the maximum score exceeds the threshold.
\begin{equation}
    s^*\mathop{\lessgtr}_{in-set}^{OOS}\theta \quad\mathrm{where}\quad s^* = \max_{i} \left\{s_i\right\}
\label{eq:1}
\end{equation}

In this study, we do not consider the complementary task of closed-set speaker identification, as our preliminary analysis shows that once it is determined that a speech sample belongs to a watchlist speaker, the systems considered in this study can determine the speaker's identity with a high precision~($>99\%$).

We consider several evaluation scenarios with different watchlist sizes and speech durations. As mentioned in Section \ref{sec:intro}, the false alarm rate increases as the watchlist size grows. 
Speech duration also plays a role in the detection performance, with longer durations typically expected to provide more reliable speaker representations, hence providing better speaker recognition performance \cite{sadjadi20192018}. Therefore, we design experiments that explore different enrollment and test durations for this task. 

\vspace{-0.2cm}

\subsection{Metrics}
\label{ssec:metrics}

The watchlist-based speaker detection task is prone to two types of errors: false rejects where an in-set speaker is incorrectly detected as not belonging to watchlist, and false alarms where an OOS speaker is incorrectly detected as belonging to watchlist. The false alarm rate~(FAR) and false rejection rate~(FRR) are determined by setting a threshold on the speaker comparison scores. We employ the commonly used equal error rate~(EER) as the primary performance metric. The EER captures the performance at the operating point where FAR equals FRR. However, practical systems seldom operate at this operating point, and thresholds are generally selected to suit the business needs. Therefore, we also evaluate the models at 2 other operating points, FRR@FAR=0.5\% (FRR at a threshold corresponding to 0.5\% FAR) and FAR@FRR=5\% (FAR at a threshold corresponding to 5\% FRR). 
We use a low FAR value(0.5\%) to evaluate the models at operating points typically used in practical watchlist-based speaker detection services \cite{singer2004multi}.
\vspace{-0.2cm}

\subsection{Performance improvements}
\label{ssec:improve}
In addition to evaluating existing systems for the watchlist detection task, we explore $3$ different techniques that have been shown in the literature to improve speaker verification performance to understand if the performance gains also translate to better performance on the OSI task.\\
\textbf{Score normalization:}
Score normalization has been previously shown to improve OSI performance with GMM-based systems \cite{fortuna2004relative,font2029mce}. We conduct experiments to determine whether score normalization techniques are as effective when used along with neural network based speaker modeling approaches. 
In particular, we use the adaptive score normalization(AS-Norm) technique \cite{cumani2011comparison} that is currently considered the most effective score normalization method for speaker recognition. \\
\textbf{Score calibration:}
Quality metric function~(QMF) based score calibration has shown to improve speaker recognition performance \cite{garcia2004use}. In particular, utterance duration, embedding magnitude, average of scores over an impostor cohort set, and signal-to-noise ratio(SNR) have been used as quality measures to calibrate the scores to model the different variabilities that could be encountered, and to suppress the dependence of the scores on these variabilities \cite{thienpondt2021idlab, hasan2013crss}. 
Overall, we train a logistic regression based calibrator that performs an affine transformation on the cosine similarity scores according to (\ref{eq:qmf}), where $s$ denotes the cosine similarity score weighted by $w_0$, $q_k$ refers to the value of the different quality measures weighted by $w_k$ and $b$ is the bias. Here, $K$ is the number of variability factors that are used in the calibration function.

\begin{equation}
    q(s) = w_{0}*s + \sum_{k=1}^K w_k * q_k + b
\label{eq:qmf}
\end{equation}
\textbf{Score fusion:}
In speaker recognition, the score-level fusion of multiple systems
is a well established technique used to improve performance \cite{huh2023voxsrc}. It works by leveraging complementary information captured by different speaker embedding models. This is realized through combining the output from these systems into a single score to be used for subsequent tasks.
Our goal is to quantify the performance gains one can obtain for OSI through a simple score fusion technique that uses an unweighted average of scores from different systems.
\vspace{-9pt}

\section{Baseline systems}
\label{sec:baseline}
In this work, we consider three baseline systems for experiments including two ResNet based systems, as well as an ECAPA-TDNN ($C=512$) based system that uses the WavLM-Large~\cite{chen2021wavlm} as the upstream front-end.\\
\textbf{ResNet-34}: The first ResNet based system is based on a publicly available model\footnote{\url{https://github.com/clovaai/voxceleb_trainer}} to facilitate reproducible experiments for the research community. The model is a variant of ResNet34 with half the number of channels in each residual block. It uses 64-dimensional log mel-spectrogram features as input to extract 512-dimensional speaker embeddings from variable duration speech by aggregating temporal frames through attentive statistics pooling (ASP)~\cite{okabe2018attentive}. It is trained using the Adam optimizer along with the angular prototypical loss~\cite{chung20b_interspeech}.\\
\textbf{ResNet-124}: The second, larger, ResNet based system adopts a modified architecture ~\cite{garcia2020magneto} with 128 base channels and 124 ($\{6, 16, 36, 3\}$) layers. It uses 80-dimensional log mel-spectrogram features as input to extract 256-dimensional speaker embeddings by pooling frame-level statistics (i.e., the mean and standard-deviation) across the duration of the input. 
It is trained using the stochastic gradient descent (SGD) optimizer with momentum (0.9) and the additive angular margin (AAM) loss. The training process comprises two stages, with 2-second chunks and a margin parameter of $0.2$ in the first stage, and 6-second chunks and a margin value of $0.5$ in the fine-tuning stage.\\
\textbf{WavLM-TDNN}: This model is also trained using SGD with momentum (0.9) and AAM loss. A weighted sum of frame-level hidden state representations from WavLM are used as input features for the TDNN. The training process consists of three stages; in the first stage, WavLM parameters remain frozen, and the TDNN model is trained using 2-3 second long segments and a loss margin of $0.2$. In the second stage, the TDNN remains frozen, while WavLM (except for the convolutional encoder) is fine-tuned using 2-3 second chunks, a margin value of $0.2$, and a small learning rate. Finally, in the third stage, the combined model is jointly optimized on 6-second chunks with a margin value of $0.4$. 

All models are trained on Vox2-dev set with data augmentation using noise samples and room impulse responses (RIR) from MUSAN~\cite{musan} and RIR~\cite{rir} datasets, respectively. For the larger ResNet and WavLM-TDNN models, we also include speed perturbed copies of the training data with ratios of 0.9 and 1.1 to increase the speaker diversity. However, the speed perturbed copies are not used in the large-margin fine-tuning stage for either of the systems.
The models used in this work obtain a competitive speaker verification performance. On the VoxCeleb-1 original verification task (Vox1-O), the ResNet-34, ResNet-124 and WavLM-TDNN models achieve an EER of 1.02\%, 0.41\% and 0.52\% respectively. Note that the ResNet34 model is adopted \textit{as is} without fine-tuning with a larger margin and using longer segments.

\begin{figure}[!t]
     \centering
     \begin{subfigure}[b]{0.47\textwidth}
         \centering
         \includegraphics[scale=0.33]{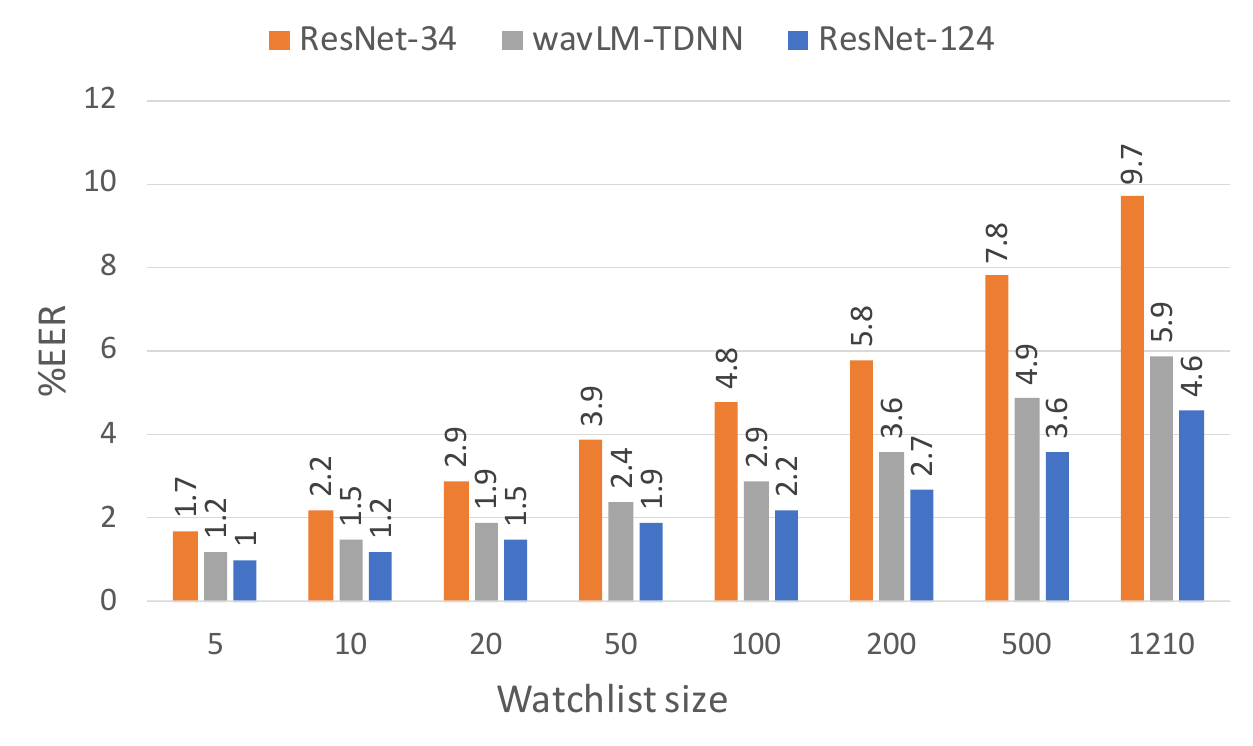}
         \caption{}
         \label{fig:ws_size_EER}
     \end{subfigure}
     \begin{subfigure}[b]{0.47\textwidth}
         \centering
         \includegraphics[scale=0.33]{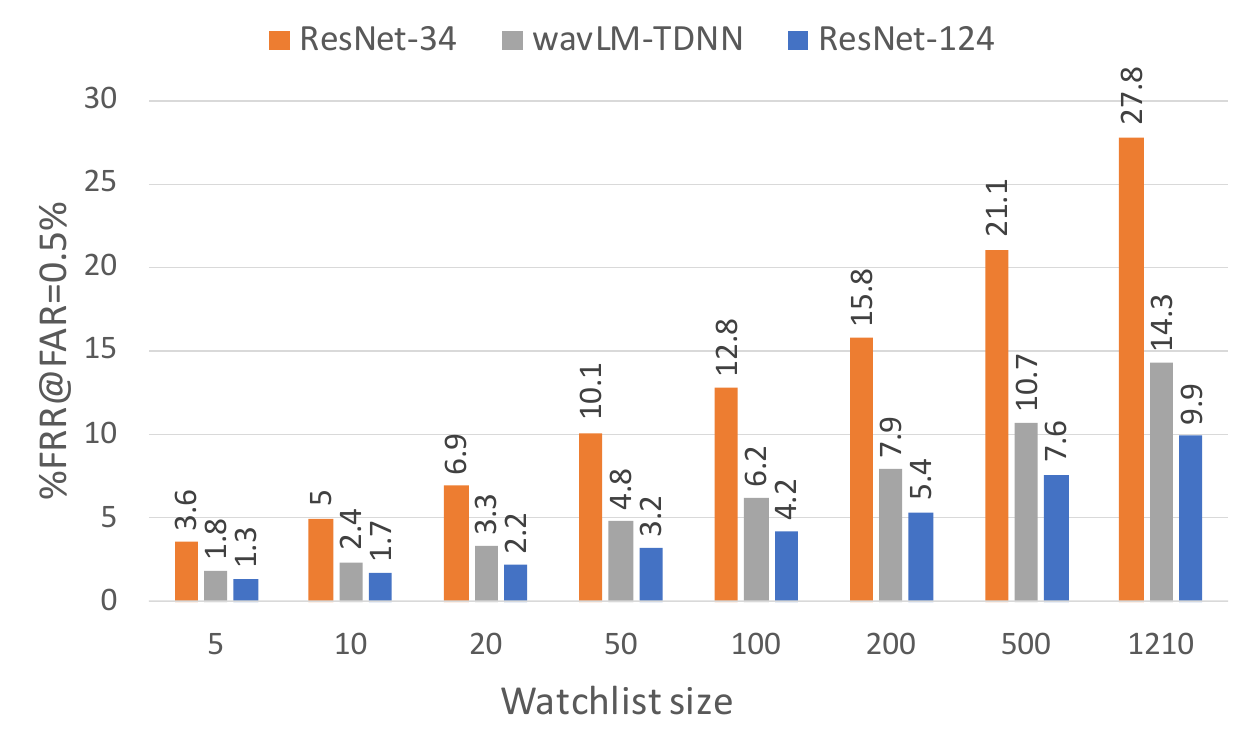}
         \caption{}
         \label{fig:ws_size_FRR}
     \end{subfigure}
     \begin{subfigure}[b]{0.47\textwidth}
         \centering
         \includegraphics[scale=0.33]{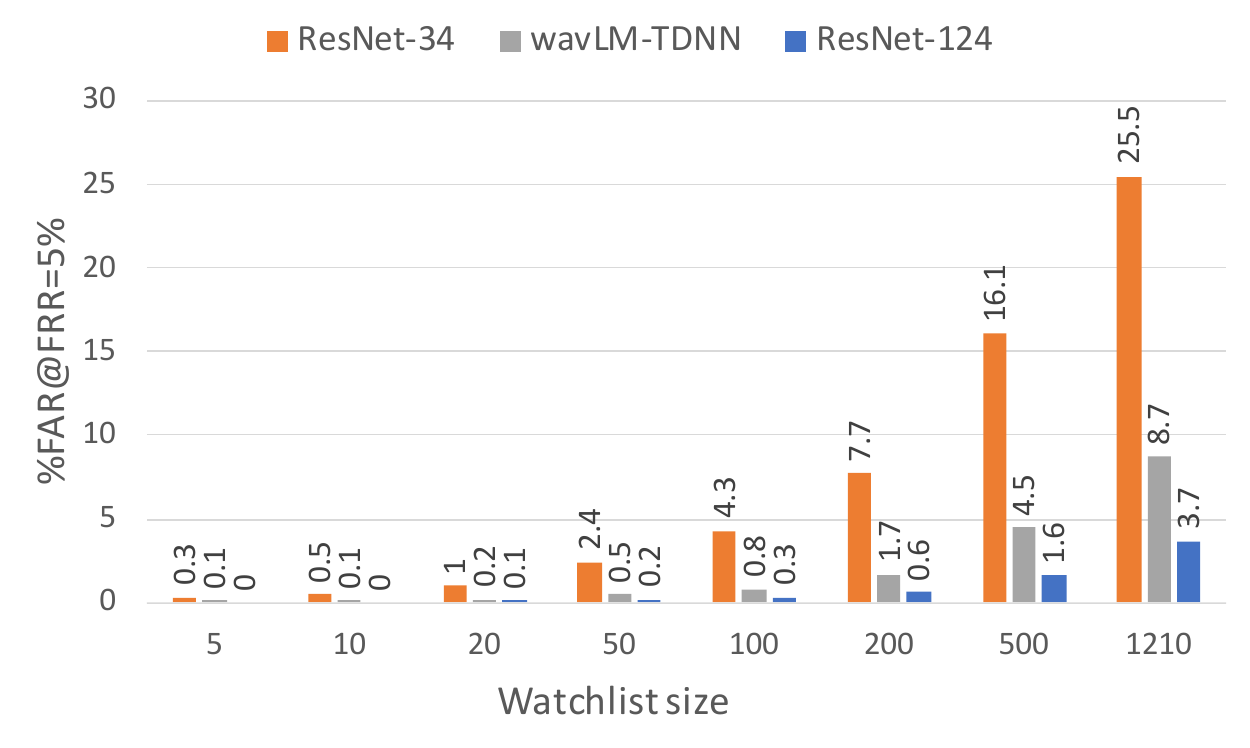}
         \caption{}
         \label{fig:ws_size_FAR}
     \end{subfigure}
        \caption{Effect of watchlist size on the detection performance of $3$ different models evaluated using $3$ metrics characterizing different performance operating points.}
        \label{fig:ws_size}
        \vspace{-0.6cm}
\end{figure}

For speaker pair comparisons, the cosine similarity score between the corresponding embedidng vectors is used. We also implement an adaptive score normalization scheme~\cite{cumani2011comparison} to reduce the within trial score variability by normalizing per trial score using the statistics calculated with the top-$k$ ($k=1000$) cohorts. 
\vspace{-9pt}
\vspace{-0.2cm}

\begin{figure*}[!htb]
     \centering
     \begin{subfigure}[b]{0.3\textwidth}
         \centering
         \includegraphics[trim={5mm 0mm 20mm 32mm},clip,width=\textwidth]{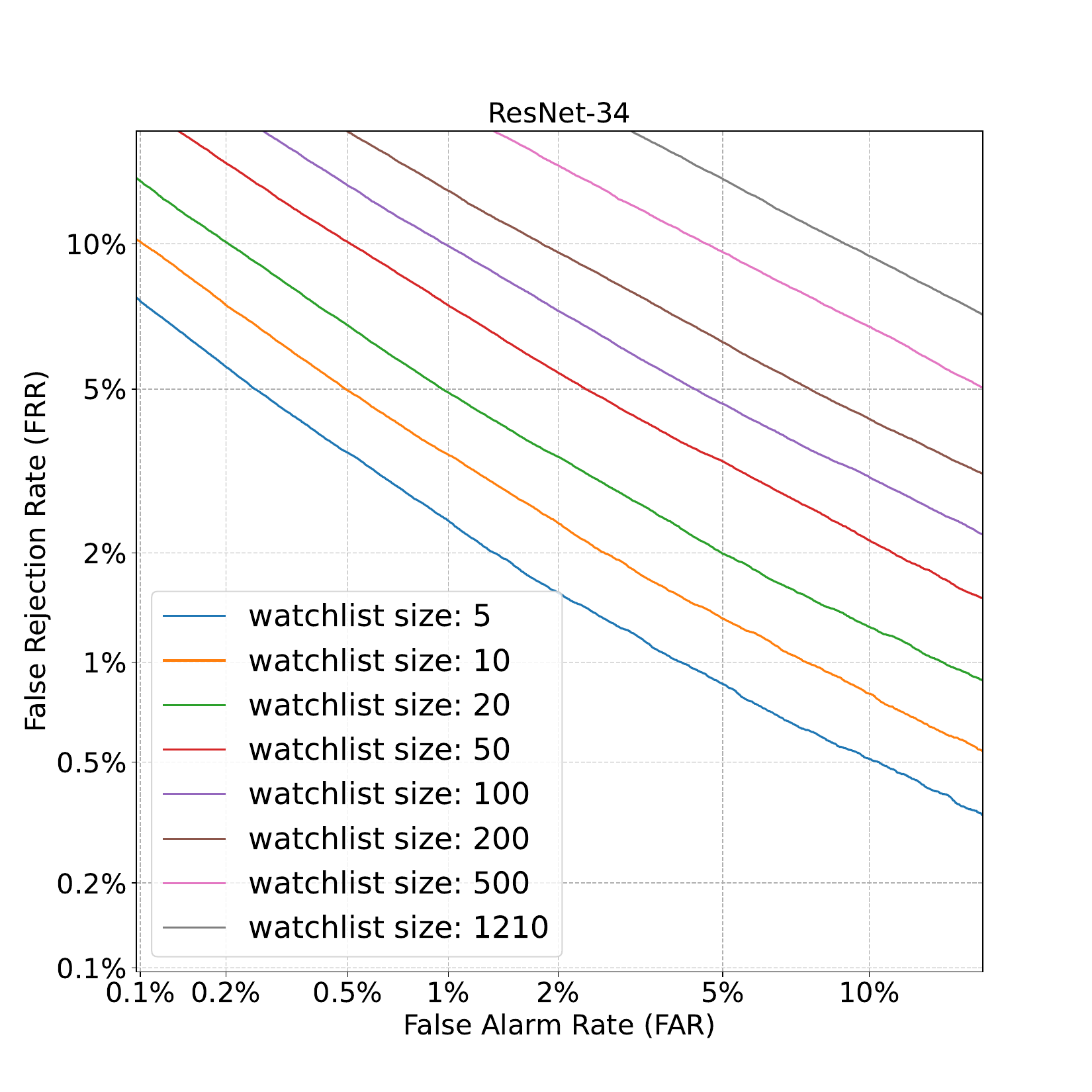}
         \caption{Effect of watchlist size with 30s enroll and 5s test duration.}
         \vspace{-0.2cm}
         \label{fig:ws_size_det}
     \end{subfigure}
     \hfill
     \begin{subfigure}[b]{0.3\textwidth}
         \centering
         \includegraphics[trim={5mm 0mm 20mm 32mm},clip,width=\textwidth]{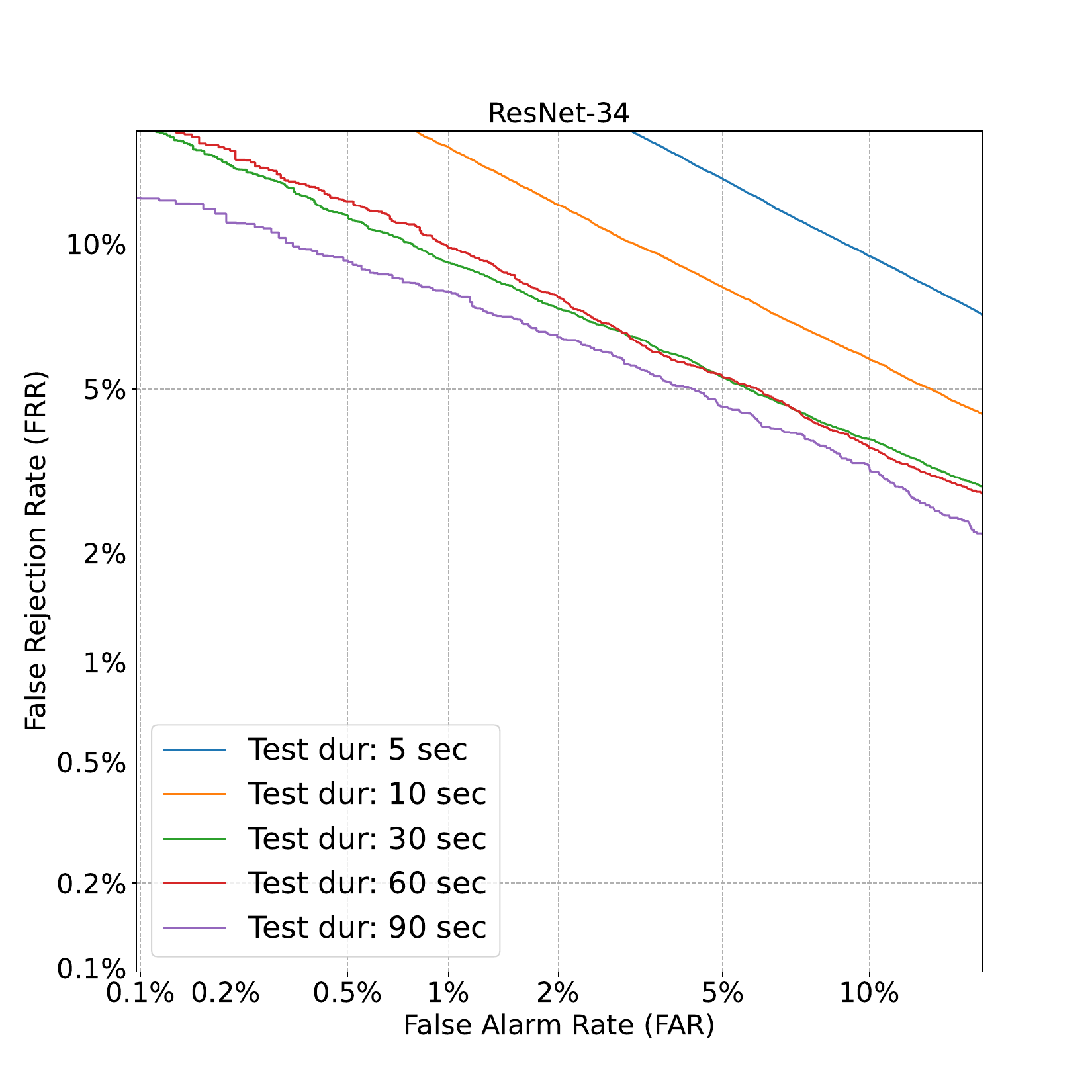}
         \caption{Effect of test duration for 30s enroll and 1210 watchlist size}
         \vspace{-0.2cm}
         \label{fig:test_dur}
     \end{subfigure}
     \hfill
     \begin{subfigure}[b]{0.3\textwidth}
         \centering
         \includegraphics[trim={5mm 0mm 20mm 32mm},clip,width=\textwidth]{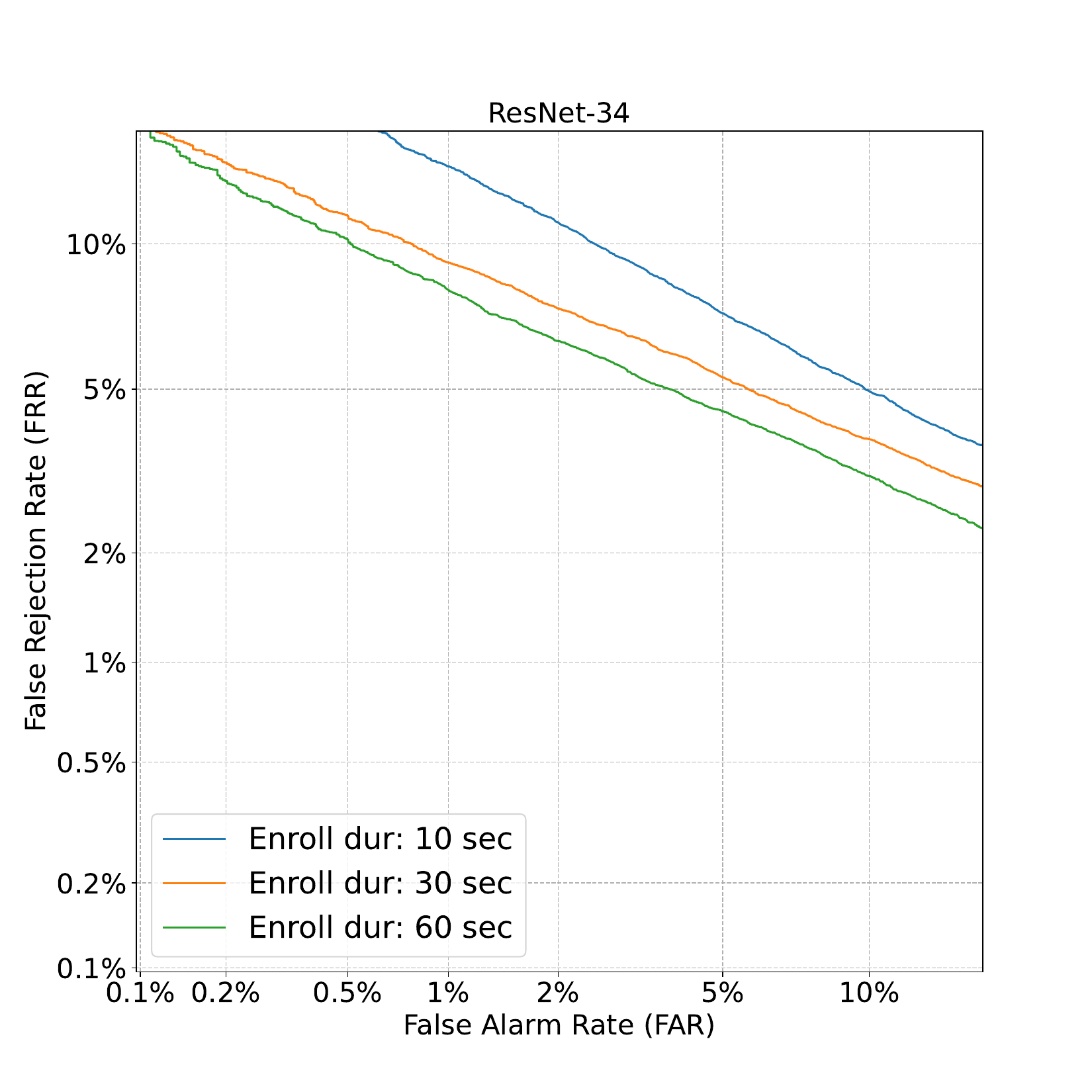}
         \caption{Effect of enrollment duration for 30s test and 1210 watchlist size}
         \vspace{-0.2cm}
         \label{fig:enroll_dur}
     \end{subfigure}
        \caption{Watchlist detection DET curves of the ResNet-34 model for different (a) watchlist sizes, (b) test durations and (c) enrollment durations.
        As the watchlist size increases, the curves shift away from origin denoting poorer detection performance. The FAR values increase more prominently than FRR.
        The performance improves as the enroll/test duration increases, but with diminishing returns.}
        \vspace{-0.5cm}
        \label{fig:det_var}
\end{figure*}

\section{Experiments and Results}
\vspace{-0.2cm}
We use one enrollment segment for each in-set speaker, and the remaining segments for that speaker are used as target samples. All segments from the OOS speakers serve as non-target samples. For each experiment, we select an enrollment and test duration which is kept fixed for that experiment. We extract speaker embeddings from the first \textit{t} seconds of each recording (i.e., we use the audio corresponding to full videos in VoxCeleb and no VAD is applied).
Unless mentioned otherwise, we use fixed enrollment and test speech duration of $30$ seconds and $5$ seconds, respectively.
\vspace{-0.3cm}
\subsection{Effect of watchlist size}
\vspace{-0.2cm}
\label{ssec:ws_size}
Figure \ref{fig:ws_size} presents the speaker detection performance of the three models detailed in Section \ref{sec:baseline} in terms of EER, FRR@FAR=0.5\% and FAR@FRR=5\% as discussed in Section \ref{ssec:metrics} for different watchlist sizes.
The performance, as measured by all $3$ metrics, clearly degrades with increasing watchlist size. In particular, the FAR measurably increases for all $3$ models. For instance, the FAR@FRR=5\% of the ResNet-34 model increases by almost $7$ times (2.4\%\textrightarrow16.1\%)  as the watchlist size grows from $50$ to $500$. For the same experiment, the FRR@FAR=0.5\% is more than doubled (10.1\%\textrightarrow21.1\%).

Figure~\ref{fig:ws_size_det} shows the detection error trade-off~(DET) curves for the ResNet-34 model for different watchlist sizes.
It is clear that the performance degradation with increasing watchlist size is consistent across all measured operating points of the system.

\vspace{-0.3cm}
\subsection{Effect of enrollment/test duration}
Figures \ref{fig:test_dur} and \ref{fig:enroll_dur} present the DET plots with varying test and enrollment speech durations, respectively, using a fixed watchlist size of $1210$. There is an evident performance gain as the speech duration (both enrollment and test) increases. In particular, an enrollment/test duration of $30$ seconds seems to provide decent performance, with further increase in duration only leading to marginal improvements. 
These results highlight a trend of diminishing returns with longer speech durations (consistent with the findings in \cite{sadjadi20192018}), and can aid the design of practical systems.
\vspace{-0.3cm}
\label{ssec:res_duration}

\begin{figure}[!t]
  \centering
  \includegraphics[trim={5mm 0mm 20mm 32mm},clip,width=0.7\linewidth]{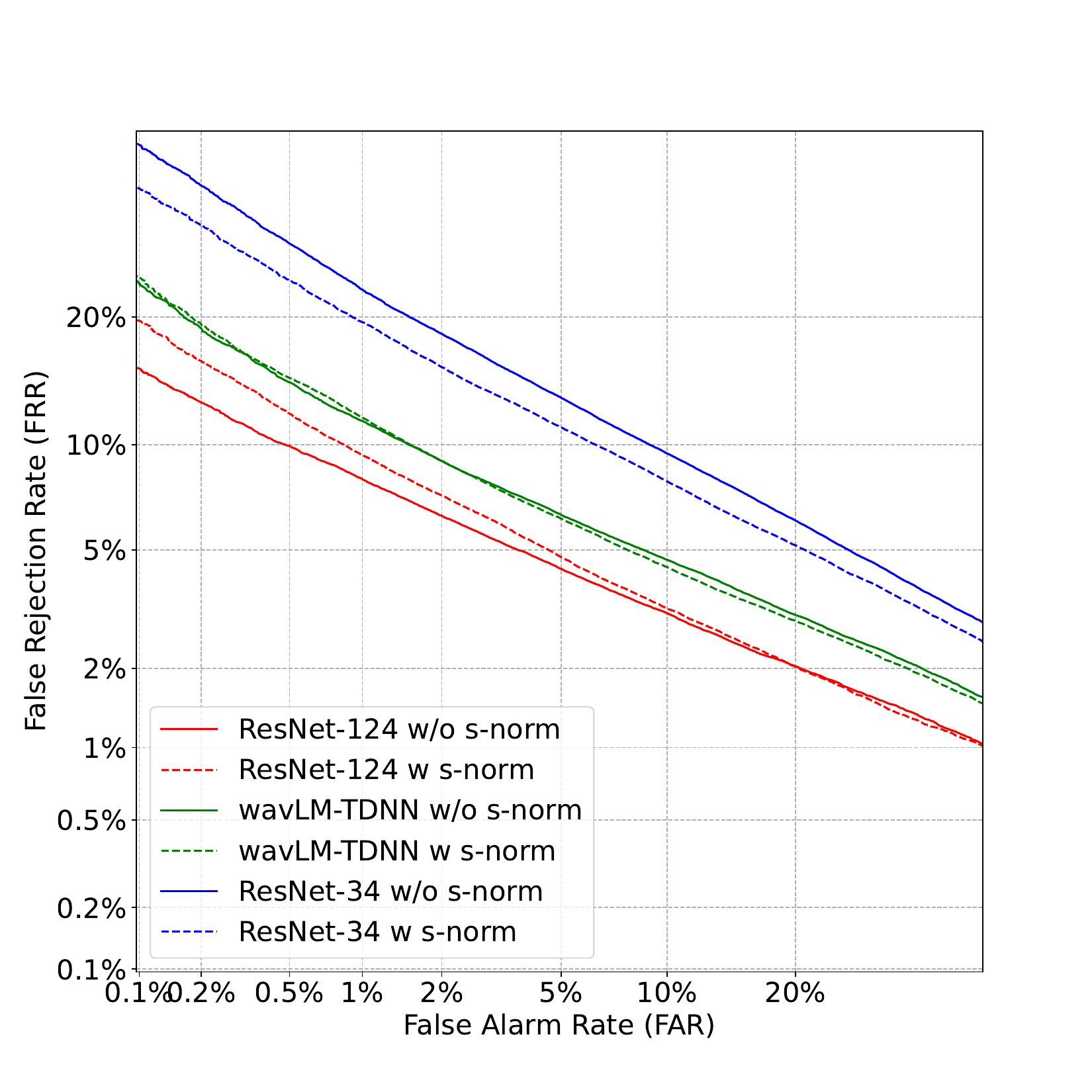}
  \caption{DET curves of different models with and without score normalization (dashed and solid lines respectively) for watchlist size of 1210. The weaker ResNet-34 model clearly benefits from s-norm across all operating regions. The benefit of s-norm on the stronger models (ResNet-124 and WavLM-TDNN) can be found at the low FRR operating regions.}
  \vspace{-0.5cm}
  \label{fig:snorm_det}
\end{figure}

\subsection{Effect of score normalization}
\label{ssec:score_norm}
Figure \ref{fig:snorm_det} shows the DET plots (watchlist size of $1210$) of the $3$ models with and without the adaptive score normalization denoted by dashed and solid lines, respectively. The score normalization has an evident benefit on the weaker performing model (ResNet-34), where the performance improves across all the operating points considered. The effect of the score normalization is less clear on the stronger models (ResNet-124 and WavLM-TDNN).
For these models, the score normalization degrades the performance across large operating regions, with improved performance in the low FRR region. These results are in contrast to previously published findings where the score normalization was shown to result in a definitive performance gain \cite{fortuna2004relative,font2029mce}. 
We hypothesize that this could be due to the traditional, less performant systems and smaller size of experiments that have been studied so far. This further highlights the importance of revisiting the watchlist-based speaker detection problem through the lens of the more recent neural-network based systems.
\vspace{-0.1cm}
\vspace{-0.2cm}
\subsection{Effect of score calibration}
\begin{figure}[ht!]
     \centering
     \begin{subfigure}[b]{0.5\textwidth}
         \centering
        \includegraphics[width=0.75\columnwidth]{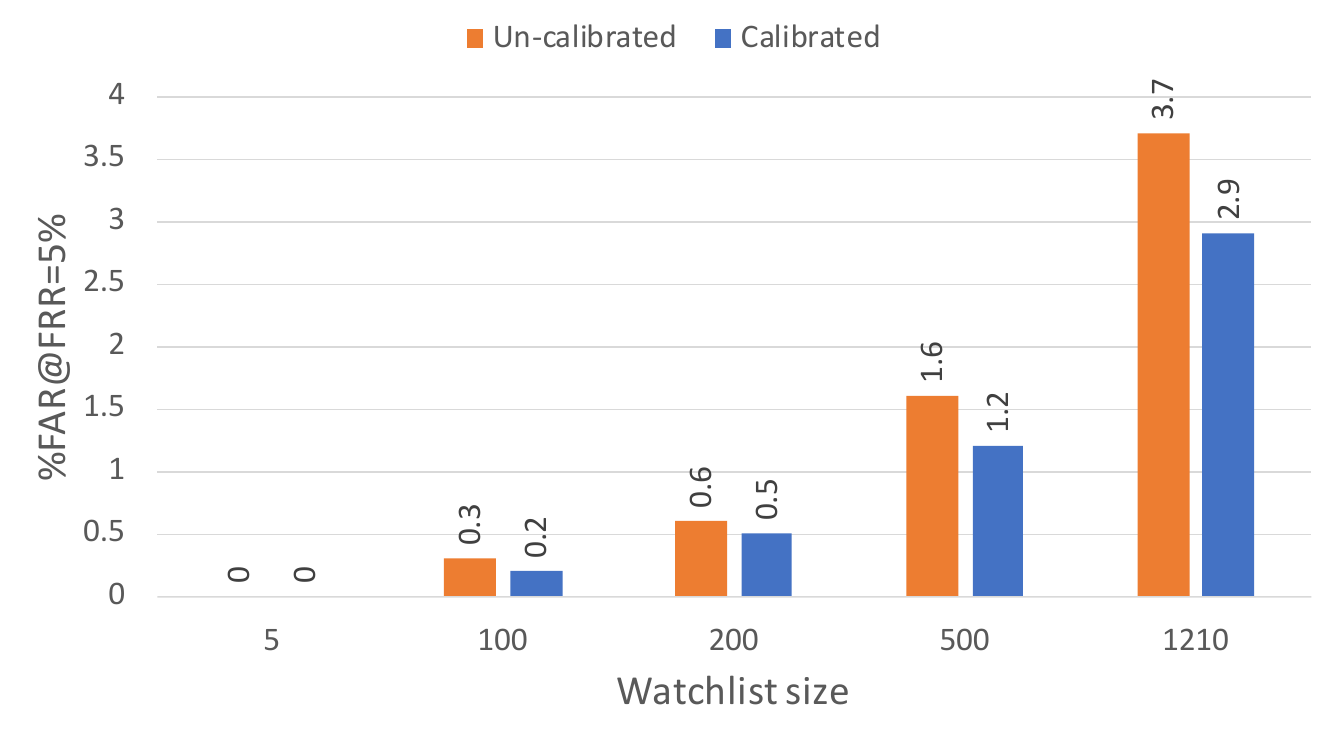}
         \caption{Score calibration performs better than without calibration.}
         \label{fig:qmf}
     \end{subfigure}
     \begin{subfigure}[b]{0.5\textwidth}
         \centering
         \includegraphics[width=0.75\columnwidth]{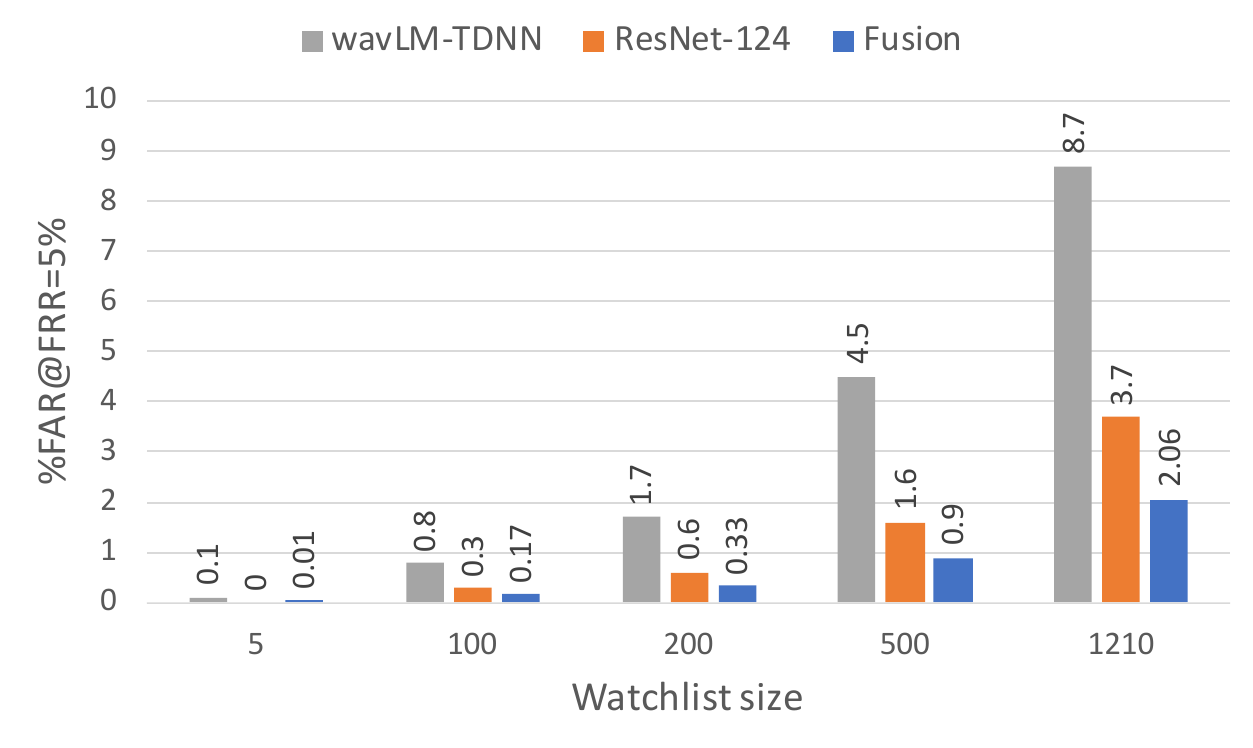}
         \caption{Score fusion outperforms individual systems for all watchlist sizes.}
         \label{fig:fuse_far}
         \vspace{-0.2cm}
     \end{subfigure}
        \caption{Watchlist detection performance(\%FAR@FRR=5\%) of (a) QMF score calibration and (b) score fusion of 2 systems.}
        \vspace{-0.6cm}
        
\end{figure}

\vspace{-0.2cm}
\label{ssec:score_calibration}
As mentioned in Section \ref{ssec:improve}, we train a linear calibrator to model and compensate for the different variabilities on a per trial basis. In particular, we consider the \emph{max} and \emph{min} of embedding magnitude, and average of impostor scores following \cite{thienpondt2021idlab}, as well as SNR.
We train the calibrator on the Vox2-dev dataset using the logistic regression according to (\ref{eq:qmf}) with $K=6$ on the scores obtained through the ResNet-124 model.
We compare the performance of the calibrated and uncalibrated systems for different watchlist sizes in Figure \ref{fig:qmf}. It can be observed that the calibrated system achieves better performance than without calibration across all watchlist sizes. In particular, the FAR@FRR=5\% metric improves by relative 21\% on the largest watchlist size setting (3.7\%\textrightarrow2.9\%)

\vspace{-0.2cm}
\subsection{Effect of score fusion}
\vspace{-0.2cm}
\label{ssec:score_fusion}
Figure \ref{fig:fuse_far} presents the \%FAR@FRR=5\% for the individual systems (wavLM-TDNN, ResNet-124) as well as their fusion at a range of different watchlist sizes. Clearly, a simple score fusion can achieve significantly better performance than the best individual system. For example, at a watchlist size of $1210$, the fusion improves the FAR compared to the best single system by relative 44\%.
This confirms that the benefits observed through the score fusion on speaker recognition task also translate to the task of watchlist detection. 
However, note that the degradation in FAR as the watchlist size increases is still similar to that of the single systems, highlighting the need for specialized techniques to alleviate the ``false alarm problem" in OSI.
\vspace{-0.5cm}
\section{Conclusions}
\vspace{-9pt}
We presented the first public benchmark setup to facilitate systematic evaluation of the open-set speaker identification tasks. Through extensive experiments with three strong neural-network based speaker embedding models, we studied the effect of watchlist size and speech duration on OSI performance. We confirmed that increasing the watchlist size results in substantial degradation in false alarm rates, and that this is still a challenging, unsolved problem in the speaker recognition domain. We further showed that adaptive score normalization, a commonly adopted technique to improve speaker verification performance, improves OSI performance only for models with weaker speaker discrimination power. Score calibration and fusion were found to consistently provide significant gains in OSI as well.
Through this open evaluation framework, we would like to draw the attention of the research community, and hope to foster innovations in tackling this important, yet under-explored, problem. 
Future work will focus on other score normalization and calibration techniques to improve the detection performance across different watchlist sizes.

\bibliographystyle{IEEEtran}
\bibliography{mybib}

\end{document}